 \documentclass[manuscript]{emulateapj}

 \usepackage{apjfonts}

\slugcomment{to appear in the Astrophysical Journal, Letters}

\shorttitle{Type I\lowercase{a} Supernova Progenitors}
\shortauthors{Hachisu et al.}

\begin{document}

\title{Supersoft X-ray Phase of Single Degenerate 
Type I\lowercase{a} Supernova Progenitors in Early Type Galaxies}

\author{Izumi Hachisu}
\affil{Department of Earth Science and Astronomy,
College of Arts and Sciences, University of Tokyo,
Komaba 3-8-1, Meguro-ku, Tokyo 153-8902, Japan}
\email{hachisu@ea.c.u-tokyo.ac.jp}

\author{Mariko Kato}
\affil{Department of Astronomy, Keio University,
Hiyoshi 4-1-1, Kouhoku-ku, Yokohama 223-8521, Japan}
\email{mariko@educ.cc.keio.ac.jp}

\and

\author{Ken'ichi Nomoto}
\affil{Institute for the Physics and Mathematics of the Universe,
University of Tokyo, Kashiwanoha 5-1-5, Kashiwa, Chiba 277-8583,
Japan}
\email{nomoto@astron.s.u-tokyo.ac.jp}

\begin{abstract}
In the single degenerate (SD) scenario for Type Ia supernova (SN~Ia)
progenitors, an accreting white dwarf (WD) is expected to undergo a
supersoft X-ray source (SSS) phase.  Recently, 
\citet[][hereafter GB10]{gil10} claimed that observed
X-ray fluxes of early type galaxies would be too low to be
consistent with the prediction of the SD scenario based on rather
simple assumptions.  We present realistic evolutionary models of SD
systems and calculate durations of SSS phases.  In most cases,
accreting WDs spend a large fraction of time in the optically thick
wind phase and the recurrent nova phase rather than the SSS phase.
Thus the SSS phase lasts only for a few hundred thousand years.  This
is by a factor of $\sim 10$ shorter than those adopted by GB10
where the SN~Ia progenitor WD was assumed to spend most of its life
as a SSS.  The theoretical X-ray luminosity of the SSS
has a large uncertainty because of the
uncertain atmospheric model of mass-accreting WDs and absorption of
soft X-rays by the companion star's cool wind material.  We thus adopt
an average of the observed fluxes of existing symbiotic SSSs, i.e.,
$\sim 0.4\times 10^{36}$~erg~s$^{-1}$ for 0.3--0.7 keV.
Using these SSS duration and soft X-ray luminosity,
we show that the observed X-ray flux obtained by GB10 is rather
consistent with our estimated flux in early type galaxies based
on the SD scenario.
This is a strong support for the SD scenario as a main-contributor
of SNe~Ia in early type galaxies.
\end{abstract}

\keywords{binaries: close --- galaxies: evolution
 --- stars: winds, outflows --- supernovae: general --- X-rays: binaries}

\section{Introduction}
Type Ia supernovae (SNe~Ia) play very important roles in astrophysics
as a standard candle to measure cosmological distances as well as
the production site of a large part of iron group elements.
However, the nature of SN~Ia progenitors has not been
clarified yet \citep[e.g.,][]{hil00, liv00, nom97, nom00}.
It has been commonly agreed that the exploding star
is a carbon-oxygen (C+O) white dwarf (WD) and the observed
features of SNe~Ia are better explained by the Chandrasekhar mass
model than the sub-Chandrasekhar mass model.
However, there has been no clear observational indication
as to how the WD mass gets close enough to the Chandrasekhar mass for
carbon ignition \citep[$M_{\rm Ia}=1.38~M_\sun$ in][]{nom82},
i.e., whether the WD accretes H/He-rich matter from
its binary companion [single degenerate (SD) scenario] or two C+O WDs
merge [double degenerate (DD) scenario].

The X-ray signature of these two possible paths are very different.
It is believed that no strong X-ray emission is expected from 
the merger scenario until shortly before the SN~Ia explosion. 
On the other hand,
the accreting WD becomes a supersoft X-ray source (SSS) long before
the SN~Ia explosion, for a million years.  In order to constrain
progenitor models in early type galaxies,
\citet[][hereafter, GB10]{gil10} recently
obtained the $0.3-0.7$ keV soft X-ray luminosity $L_{\rm X,obs}$
and the $K$-band luminosity $L_K$ for several early type galaxies.
They compared $L_{\rm X,obs}$ with those predicted from the SN~Ia birth
rate estimated from $L_K$.  Their predicted
X-ray luminosity $L_{\rm X,SSS}$ from the SD scenario is 40--70 times
larger than $L_{\rm X,obs}$ and they concluded that no more than
five percent of SNe~Ia in early type galaxies can be produced
by mass-accreting WDs of the SD scenario.  However, their 
$L_{\rm X,SSS}$ is based on the following assumptions, which 
involves large uncertainties.  They assumed 
that (1) all the accreting WDs are in the SSS phase 
which typically lasts for two million years before a SN~Ia
explosion and (2) the observed (including absorption) 0.3--0.7 keV
soft X-ray flux is as large as 
$(3-5)\times 10^{36}$~erg~s$^{-1}$ per source.

In a canonical SD scenario \citep[e.g.,][]{hkn99}, however, the
accreting WDs usually spend a large fraction of the lifetime in the
optically thick wind phase and in the recurrent nova phase, so that
a duration of the SSS phase is much shorter than that assumed by GB10.
Moreover, the X-ray luminosity of the symbiotic SSS has a large
uncertainty because of the uncertain atmospheric model of
mass-accreting WDs and absorption of soft X-rays by the companion
star's cool wind material.  In these situations, it is reasonable
that we use observed fluxes of existing symbiotic SSSs.

In this Letter, we show that the soft X-ray fluxes observed in
early type galaxies are consistent with those expected from the SD 
model, if we adopt a realistic scenario of binary evolutions
including a shorter SSS duration and a much more absorbed soft X-ray
flux estimated from existing symbiotic SSSs.  We first show the short
SSS duration of our SD models in Section 2.  In Section 3,
we estimate absorbed soft X-ray fluxes in  early type galaxies.
Discussion follows in Section 4.


\begin{figure}
\epsscale{1.15}
\plotone{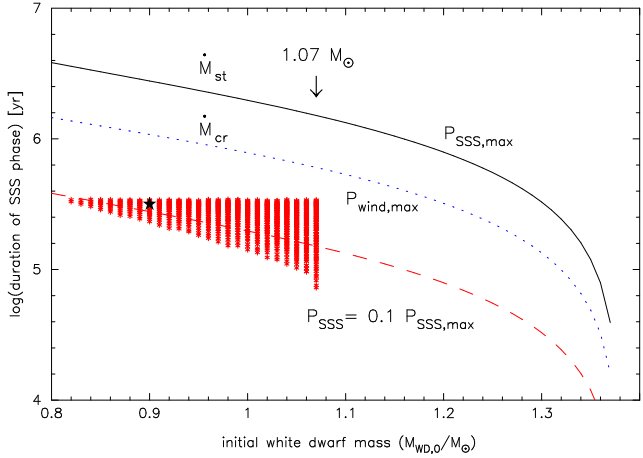}
\caption{
Duration of a SSS phase against the initial WD mass for the WD+RG
systems.  Red asterisks denote the SSS durations of our progenitor
models that produced a SN~Ia.  {\it Solid}: $P_{\rm SSS, max}$,
the maximum duration of the SSS phase calculated from Equation
(\ref{maximum_duration_sss}).  {\it Dotted}: $P_{\rm wind, max}$,
the maximum duration of the wind phase calculated
from Equation (\ref{maximum_duration_wind}).
{\it Dashed}: $P_{\rm SSS}=0.1 P_{\rm SSS,max}$.
{\it Star symbol}:
a model corresponding to Figure  \ref{evl_strip_wd090m130p400}.
\label{sss_duration}}
\end{figure}

\section{Duration of Supersoft X-ray Source Phase}
In early type galaxies, star formation virtually stopped several Gyr
ago. Therefore, only the white dwarf (WD) + red giant (RG) binary 
systems with the initial RG companion mass of
$M_{2,0}\lesssim 1.3~M_\sun$ 
\citep[e.g.,][]{hkn99} contribute to SNe~Ia.  For the white dwarf (WD)
+ main-sequence (MS) star binary systems, more massive companions
\citep[usually $M_{2,0}\gtrsim 1.8~M_\sun$, see, e.g.,][]{hknu99}
are required but there remain no such massive ones because they have
already evolved off.  Therefore, here we estimate
the duration of a supersoft X-ray phase only for the WD+RG systems.

Our evolutionary model for the SN~Ia progenitor is as follows:
The binary evolution starts from zero-age main-sequence star pairs
with a given set of the primary mass ($M_{1,i}$),
secondary mass ($M_{2,i}$), and separation ($a_i$)
as done in \citet{hkn99}.
Unless the initial separation of the binary components is too close,
the more massive (primary) component
evolves to a RG star (with a helium core)
or an AGB star (with a C+O core) and fills its Roche lobe or
blows a superwind.  If subsequent mass transfer
from the primary to the secondary is rapid enough to form
a common envelope, the binary separation
shrinks greatly owing to mass and angular momentum losses
from the binary system during the common envelope evolution
\citep[see, e.g., Figure 1 of][for an illustration of the
binary evolution]{hkn99}.
The hydrogen-rich envelope of the primary component
is stripped away and the primary becomes either a helium star
or a C+O WD.  The helium star further evolves to a C+O WD
after a large part of helium is exhausted by core helium burning.
Thus we have a binary pair of the C+O WD and the secondary star
that is still an MS.  At this stage (denoted by $0$), the binary
becomes a pair of the C+O WD with the mass of $M_{\rm WD,0}$ and
the secondary star that is still on the MS; the mass of the
secondary star, $M_{2,0}$, is still close to $M_{2,0}\approx M_{2,i}$,
because the accreted mass during the common envelope phase
is negligibly small.  

After the secondary evolves to 
fill its Roche lobe, the WD accretes mass from the secondary and
grows to the critical mass \citep[$M_{\rm Ia}=1.38~M_\sun$ in][]{nom82}
to explode as a SN~Ia if the initial binary orbital period ($P_0$)
and the initial mass of the secondary ($M_{2,0}$) are in
the regions (labeled ``initial'') shown in Figure 1 of \citet{hkn08}.
There are two separate regions; one is
for binaries consisting of a WD and an MS (WD+MS)
and the other is binaries consisting
of a WD and a RG (WD+RG).  In this figure,
the metallicity and the initial mass of the WD were assumed
to be $Z=0.02$ and $M_{\rm WD,0}=1.0~M_\sun$.
In early type galaxies considered here, however, MS
stars more massive than $1.3~M_\sun$ have already evolved off
to RGs or WDs, so that only the WD+RG systems can produce
SNe~Ia.

After the mass transfer begins from the RG to the WD in our 
WD+RG systems, a steady-state supersoft X-ray source (SSS)
phase can be realized only when the mass accretion rate onto the WD
satisfies
\begin{equation}
\dot M_{\rm st}<\dot M_{\rm WD}<\dot M_{\rm cr},
\label{steady_sss_condition}
\end{equation}
where $\dot M_{\rm st}$ is the lower limit mass accretion rate for stable
hydrogen shell burning on the WD, $\dot M_{\rm WD}$ the mass accretion
rate onto the WD, and $\dot M_{\rm cr}$ the critical mass accretion rate
above which the WD envelope expands to blow optically thick winds
\citep{hkn96, hkn99, hknu99}.
\citet{nom07} obtained these critical rates as
\begin{equation}
\dot M_{\rm st}=3.066\times 10^{-7}\left({{M_{\rm WD}}\over 
{M_\sun}}-0.5357\right)M_\sun{\rm ~yr}^{-1},
\label{steady_state_burning}
\end{equation}
and
\begin{equation}
\dot M_{\rm cr}=6.682\times 10^{-7}\left({{M_{\rm WD}}\over 
{M_\sun}}-0.4453\right)M_\sun {\rm ~yr}^{-1}.
\label{steady_accretion_wind}
\end{equation}

Using Equation (\ref{steady_sss_condition}), we estimate
the ``maximum duration'' of the steady-state SSS phase
before the SN~Ia explosion, i.e.,
\begin{equation}
P_{\rm SSS, max}=\int_{M_{\rm WD,0}}^{M_{\rm Ia}} 
{{d M}\over{\dot M_{\rm st}}}
=3.3\times 10^{6}~\ln\left({{M_{\rm Ia}/M_\sun-0.5357}\over 
{M_{\rm WD,0}/M_\sun-0.5357}}\right){\rm ~yr}.
\label{maximum_duration_sss}
\end{equation}
We plot the value of $P_{\rm SSS,max}$ (solid line)
in Figure \ref{sss_duration}
against various initial WD masses.
If there are any paths in binary evolutions in which the mass
accretion rate would always follow Equation (\ref{steady_state_burning}),
its SSS duration would be given by this $P_{\rm SSS, max}$.
We should note that GB10 assumed essentially the same SSS duration
as this $P_{\rm SSS,max}$.  For comparison, we also plot
the maximum duration of the wind phase, $P_{\rm wind,max}$ (dotted
line), i.e.,
\begin{equation}
P_{\rm wind, max}=\int_{M_{\rm WD,0}}^{M_{\rm Ia}} 
{{d M}\over{\dot M_{\rm cr}}}
= 1.5\times 10^{6}~\ln\left({{M_{\rm Ia}/M_\sun-0.4453}\over 
{M_{\rm WD,0}/M_\sun-0.4453}}\right){\rm ~yr}.
\label{maximum_duration_wind}
\end{equation}


\begin{figure}
\epsscale{1.15}
\plotone{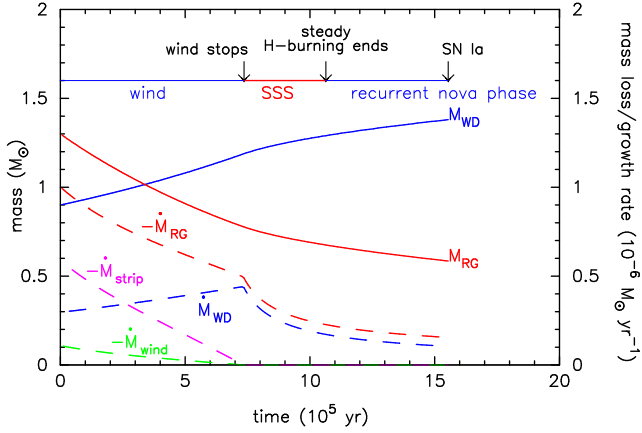}
\caption{
Time evolution of a SN~Ia progenitor system for an initial
set of parameters of 
$M_{\rm WD,0}=0.9~M_\sun$, $M_{2,0}=1.3~M_\sun$, and
$a_0=297~R_\sun$ ($P_0=400$ days).  The WD blows optically
thick winds until $t=P_{\rm wind}\sim 7.3\times 10^5$~yr,
enters the SSS phase until $\sim 10.5\times 10^5$~yr,
then becomes a recurrent nova, and finally explodes as
a SN~Ia at $t_{\rm SN~Ia}\sim 15.5\times 10^5$~yr.  We
obtain the duration of $P_{\rm SSS}\sim 3.2\times 10^5$~yr.
We include the rate of mass-stripping from the RG by the WD wind
($\dot M_{\rm strip}$, that is, 
$-\dot M_{\rm RG}\approx-\dot M_{\rm wind}-\dot M_{\rm strip}+\dot
M_{\rm WD}$ during the wind phase)
and the mass-loss during helium shell-flashes
($\dot M_{\rm WD}=\eta_{\rm H}\eta_{\rm He}|\dot M_{\rm RG}-\dot 
M_{\rm strip}|$).
See Equations (22) and (15) of \citet{hkn99}, respectively.
The WD and RG masses (solid lines) refer to the left axis
while the mass-loss/growth rates (dashed lines) refers to
the right axis.
\label{evl_strip_wd090m130p400}}
\end{figure}


\begin{figure}
\epsscale{1.15}
\plotone{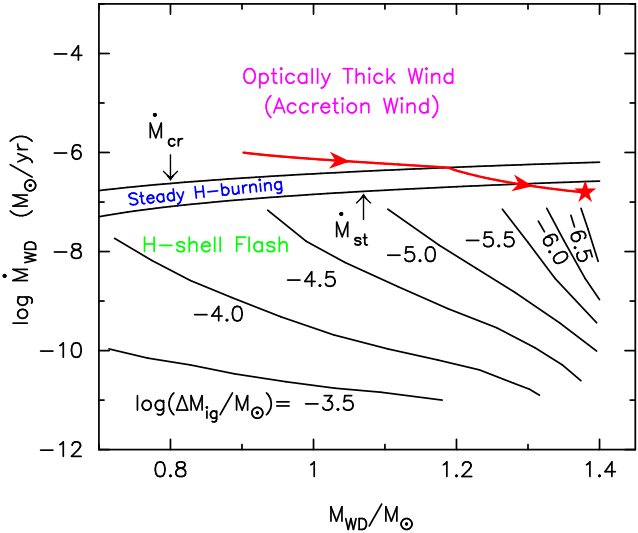}
\caption{
Evolutionary path ({\it red solid}) of a SN~Ia progenitor
(same model as in Figure \ref{evl_strip_wd090m130p400})
on the map of response of white dwarfs to mass accretion rate.
The progenitor explodes at the star mark as a SN~Ia.  
Strong optically thick winds blow above the line of
$\dot M_{\rm WD}>\dot M_{\rm cr}$.  The wind mass-loss rate is
$\dot M_{\rm wind}\approx\dot M_{\rm WD}-\dot M_{\rm cr}$.
Steady hydrogen shell burning with no optically thick winds occur
between $\dot M_{\rm st}\le\dot M_{\rm WD}\le\dot M_{\rm cr}$.
There is no steady state burning below
$\dot M_{\rm WD}<\dot M_{\rm st}$.  Instead, intermittent
shell flashes occur.  The envelope mass, $\Delta M_{\rm ig}$,
at which a hydrogen shell flash ignites, is also shown (taken
from Figure 9 of \citealp{nom82}).  Progenitor's crossing time of
the region between 
$\dot M_{\rm st}\le\dot M_{\rm WD}\le\dot M_{\rm cr}$,
i.e., the duration of a SSS phase, is relatively short.
\label{acc_map_paris2010}}
\end{figure}

The upper limit mass for C+O WDs born in binary systems was
estimated to be $\sim 1.07~M_\sun$ by \citet{ume99}.
In early type galaxies, the donor star is relatively less 
massive and cannot supply much mass to the WD.  Therefore, 
the initial mass of WDs should be as massive as $0.8~M_\sun$ or
more as shown in Figures 12 and 13 of \citet{hkn99}.
Here we adopt an upper limit of $M_{\rm WD,0}=1.07~M_\sun$
for the initial WD mass.  We have followed binary evolutions using
the same method and physical parameters as those in \citet{hkn99}.

In  Figures \ref{evl_strip_wd090m130p400} and \ref{acc_map_paris2010},
a typical example of the evolutionary path is
plotted, i.e., $M_{\rm WD,0}=0.9~M_\sun$, $M_{2,0}=1.3~M_\sun$, and
$P_0=400$ days ($a_0=297~R_\sun$).
In these figures, the evolution starts when the secondary
RG fills its Roche lobe to start mass transfer.
The mass transfer rate at the initial phase
exceeds $\dot M_{\rm cr}$, so that the WD blows optically
thick winds until $t=P_{\rm wind}\sim 7.3\times 10^5$~yr.
Then the WD enters the SSS phase until
$t\sim 10.5\times 10^5$~yr, followed by the recurrent
nova (RN) phase, and finally explodes as an SN~Ia
at $t_{\rm SN~Ia}\sim 15.5\times 10^5$~yr.
In the RN phase, the WD undergoes weak hydrogen shell-flashes
every 10 yrs or so.  We may observe a SSS phase at
every RN explosion but this duration is very short,
only 0.6 \% of the mass accretion phase, for example, in the 
RS Oph case \citep[50 days against 22 yrs; see, e.g., ][]{hac07kl}.
In this case, we have a duration of $P_{\rm SSS}\sim 3.2\times 10^5$~yr,
i.e., the crossing time of the region
between $\dot M_{\rm st}\le\dot M_{\rm WD}\le\dot M_{\rm cr}$
is $\sim 10$ times shorter than that estimated from
Equation (\ref{maximum_duration_sss}).

We have also examined $P_{\rm SSS}$ for 1239 cases with a different set
of initial parameters as shown by red asterisks in Figure \ref{sss_duration}.
The lower boundary ($P_{\rm SSS}\sim 0.7$--$3.3\times 10^5$~yr)
is mainly determined by the upper limit of 
the companion mass, i.e., $1.3~M_\sun$, while the upper boundary
($P_{\rm SSS}\sim 3.4\times 10^5$~yr)
is determined by the lowest companion mass that depends on the orbital
period as shown in Figures 12 and 13 of \citet{hkn99}.
The mean duration of these cases, which covers our progenitor regions of
SNe~Ia, is $P_{\rm SSS}\sim 2.5^{+0.9}_{-1.8}\times 10^5$~yr.
Therefore, we adopt a typical value of $P_{\rm SSS}\sim 2.5\times 10^5$~yr
in our estimate,  which is a factor of $\sim 8$ shorter than
the value assumed by GB10.

\section{Soft X-ray Flux in the SD Model}
GB10 obtained the $0.3-0.7$ keV soft X-ray to $K$-band
luminosity ratio for several early type galaxies
and compared them with the X-ray luminosity expected from 
the SN~Ia birth rate.  They assumed the number of accreting WDs
in the SSS phase as
\begin{equation}
N_{\rm WD, SSS}\approx{{\Delta M_{\rm WD}}\over{\dot M_{\rm WD}}}
 \dot N_{\rm SN~Ia}\approx P_{\rm SSS}\dot N_{\rm SN~Ia},
\label{sss_number_gilfanov}
\end{equation}
where $\Delta M_{\rm WD}$ is the mass difference between the initial
mass ($M_{\rm WD,0}$) and the final mass ($M_{\rm Ia}$),
and $\dot N_{\rm SN~Ia}$ is the SN~Ia birth rate given by
\begin{equation}
\dot N_{\rm SN~Ia} = {1\over 2}\times
3.5\times 10^{-4}\left({{L_K}\over{10^{10}L_{K,\sun}}}\right)
{\rm ~yr}^{-1}.
\label{sn1a_birth_rate_k}
\end{equation}
If we adopt our real duration of $P_{\rm SSS}\sim 2.5\times 10^5$~yr,
the number of WDs in the SSS phase is
\begin{equation}
N_{\rm WD,SSS}\approx 44\left({{L_K}\over{10^{10}L_{K,\sun}}}\right),
\label{sss_number_hkn}
\end{equation}
as tabulated in Table \ref{accreting_wds}
together with the results of GB10.
Since our real duration of $P_{\rm SSS}$ is much shorter than
those assumed by GB10, the number of accreting WDs is
by a factor of $\sim 7$ smaller than those obtained by GB10.

\placetable{accreting_wds}

We also point out that the soft X-ray flux
from mass-accreting WDs in the SSS
phase involves quite a large uncertainty.  The key point is that the
progenitor system is the WD+RG system, in which the WD
is embedded in a complex circumbinary matter.  Furthermore, we have not
yet fully understood atmospheric structures of mass-accreting WDs
in symbiotic stars \citep[e.g.,][]{jor96,ori07}.  In these situations,
it is reasonable that we use observed fluxes of symbiotic
SSSs with known distance like the member of the Small
Magellanic Cloud (SMC).  There are two well-studied symbiotic SSSs
in the SMC, SMC~3 and Lin~358.

The soft X-ray flux (0.15--2.4 keV) from SMC~3 with {\it ROSAT}
was obtained to be 
$\sim 3.3\times 10^{36}(d/65{\rm ~kpc})^2$~erg~s$^{-1}$
by \citet{kah94}.
The flux between 0.3--0.7 keV is 0.3--0.4 of the total
flux \citep[see, e.g., Figures 2--6 of][for the spectrum]{jor96},
so that $\ell_{\rm X,obs}\sim 1\times 10^{36}$~erg~s$^{-1}$ for
0.3--0.7 keV.  This Kahabka et al.'s flux has been already corrected for 
the Galactic absorption, i.e., $N_{\rm H}=3\times 10^{20}$~cm$^{-2}$
with an unknown factor, which is not specified in the paper.
Thus the absorbed flux at the Earth is smaller than
$1\times 10^{36}$~erg~s$^{-1}$.  
The 0.2--1.0 keV flux from SMC~3 outside eclipse was also 
observed with {\it XMM-Newton}, and \citet{ori07} obtained 
$\sim 3\times 10^{-12}$~erg~cm$^{-2}$~s$^{-1}$
at the Earth (private communication).
The 0.3--0.7 keV flux is about 60\% of the total flux
\citep[see Figure 2 of][for the spectrum]{ori07}, i.e.,
$f_{\rm X,obs}=1.8\times 10^{-12}$~erg~cm$^{-2}$~s$^{-1}$.
We obtain $\ell_{\rm X,obs}=4\pi d^2 f_{\rm X, obs}=7.7
\times 10^{35}(d/60{\rm ~kpc})^2~{\rm ~erg~s}^{-1}$. 
The above Kahabka et al.'s value is consistent with this flux, 
suggesting no variation of the X-ray flux during $\sim 10$~yr.
So, in the following, we use Orio et al.'s flux of 
$\ell_{\rm X,obs}=7.7\times 10^{35}{\rm ~erg~s}^{-1}$.
Adopting neutral hydrogen $N_{\rm H}=4.9\times 10^{20}$~cm$^{-2}$
and temperature $T=5.1\times 10^{5}$~K \citep[Rauch model for MOS-1 data 
in Table 1 in][]{ori07}, we have calculated the correction factor of 
the absorbed flux with different $N_{\rm H}$ but the same temperature. 
For example, the correction factor for NGC3585,
$N_{\rm H}=5.6\times 10^{20}$~cm$^{-2}$ \citep{bog10},
is calculated to be 0.88 against LMC~3 of 
$7.7\times 10^{35}{\rm ~erg~s}^{-1}$
assuming blackbody of $T=5.1\times 10^{5}$~K, 
so we have the expected absorbed flux from NGC3585
to be $\ell_{\rm X,cal}=6.8\times 10^{35}$~erg~s$^{-1}$.
In this way, we have obtained expected absorbed flux for each
early type galaxy listed in Table \ref{accreting_wds}.

Another example is Lin~358, the Small Magellanic Cloud symbiotic star.
The absorbed soft X-ray flux (0.13--1.0 keV)
from Lin~358 with {\it XMM Newton} was estimated to be
$\sim 8.3\times 10^{34}(d/ 60{\rm ~kpc})^2$~erg~s$^{-1}$ with
$N_{\rm H}=7.6\times 10^{20}$~cm$^{-2}$  and 
$T=2.3\times 10^5$~K by \citet{kah06}.
Since the spectrum is very soft,
the flux between 0.3--0.7 keV is at most 0.2 of the total
flux \citep[see Figure 3 of][for the spectrum]{ori07}.
We obtain $\ell_{\rm X,obs}\sim 1.7\times 10^{34}$~erg~s$^{-1}$ for
0.3--0.7 keV.  The converted factor to NGC3585 is 1.6 and we obtain 
the expected X-ray flux to be 
$\ell_{\rm X,cal}=2.7\times 10^{34}$~erg~s$^{-1}$.

These two symbiotic stars have very different values of X-ray flux, 
so we adopt an arithmetic mean of these two and each 
$\ell_{\rm X, cal}$ is shown in Table \ref{accreting_wds}.
Then the total expected 
supersoft X-ray luminosity of $L_{\rm X, SSS}$ is obtained as 
\begin{equation}
L_{\rm X, SSS}=N_{\rm WD,SSS}\times\ell_{\rm X,cal},
\label{total_soft_x-ray_flux}
\end{equation}
for the expected soft (0.3--0.7 keV) X-ray flux, that is also shown in 
Table \ref{accreting_wds}.  The estimated soft X-ray fluxes
are rather consistent with the observed ones of early type galaxies. 
This is in apparent contradiction with the claim made by GB10.

\section{Discussion}
Our estimated duration of $P_{\rm SSS}\sim 2.5^{+0.9}_{-1.8}\times
10^5$~yr is rather stiff as long as our SD model is concerned.
However, there is still a large uncertainty on the absorbed
soft X-ray flux.  The bright symbiotic SSSs are rather rare as
\citet{vog94} wrote ``Its (SMC~3) X-ray luminosity in the
0.1--2.4 keV energy band was estimated to be $\approx 400~L_\sun$.
This is an unusually high X-ray flux for a symbiotic binary.''
If SMC~3 is a brightest exception considering low X-ray fluxes of
other symbiotic SSSs like Lin~358, our flux estimate in 
Table \ref{accreting_wds} may be reduced further.

As one of new possible mechanisms to SNe~Ia, which
might be additional soft X-ray fluxes,
\citet{kin03} speculated that a dwarf nova could become a SSS during
the outburst if $\dot M_{\rm WD}$ temporarily increases by a factor of
$\sim 10-100$ (i.e., $\dot M_{\rm WD}>\dot M_{\rm st}$).
Such a SSS, however, cannot be realized.
Hydrogen shell-burning does not occur in such a case because
the accreted envelope mass is too small to be ignited.
For example, \citet{sta88} calculated mass accretion onto the very
massive WD of $1.35M_\sun$.  With a high accretion rate of
$1.1\times 10^{-6}M_\sun$~yr$^{-1}$, they found that hydrogen
ignites 2.6 yr after the accretion starts.  This means that
the accreted mass of $3\times 10^{-6}M_\sun$ should be required
before the hydrogen ignition.

In King et al.'s idea (see their Figure 2), the average
mass accretion rate is $1\times 10^{-9}M_\sun$~yr$^{-1}$
and the duty cycle is 0.004,
which means a burst mass accretion rate of $2.5\times 10^{-7}
M_\sun$~yr$^{-1} $ ($>\dot M_{\rm st}$).  The viscous timescale of
the accretion disk in the WD+RG systems like RS Oph is about
a few to several hundred days or so \citep[e.g.,][]{kin09}.
Therefore, the disk mass can be estimated to be 
$2.5\times 10^{-7}M_\sun$~yr$^{-1}\times (0.7-1.5)$~yr 
$\sim(2-4)\times 10^{-7}M_\sun$,
which is too small to trigger a shell burning even on a very massive WD
like in RS Oph.  Even if a disk instability occurs,
the accreted matter simply accumulates
on the WD but remains unburnt.   After a few tens of disk instabilities,
the accreted materials finally reaches the critical mass of 
$3\times 10^{-6}M_\sun$ and results in a shell flash.
This is nothing else a classical (or recurrent) nova
and its SSS duration is too short to be compared with the steady SSS
duration as mentioned earlier.

\acknowledgments
We thank Marat Gilfanov and Joanna Miko\l ajewska for useful discussion
on symbiotic SSSs like SMC~3.
We are also grateful to the anonymous referees for
their detailed comments that helped to improve the paper.
This research has been supported by 
the Grant-in-Aid for Scientific Research of the Japan Society for the
Promotion of Science (20540226, 20540227), and by World Premier
International Research Center Initiative, MEXT, Japan.

\clearpage

\begin{deluxetable}{lcrrccc}
\tablecaption{White dwarfs in supersoft X-ray source phase
\label{accreting_wds}}
\tablewidth{0pt}
\tablehead{
\colhead{galaxy} & \colhead{$L_K$\tablenotemark{a}}
& \colhead{$N_{\rm WD,SSS}$\tablenotemark{a}} 
& \colhead{$N_{\rm WD,SSS}$\tablenotemark{b}}
& \colhead{$\ell_{\rm X,cal}$\tablenotemark{c}} 
& \colhead{$L_{\rm X,obs}$\tablenotemark{a}} 
& \colhead{$L_{\rm X,SSS}$\tablenotemark{d}}  \nl
\colhead{} & \colhead{$(10^{10} L_{K, \sun})$}
& \colhead{}  & \colhead{} & \colhead{($10^{35}$~erg~s$^{-1}$)}
& \colhead{($10^{37}$~erg~s$^{-1}$)}
  & \colhead{($10^{37}$~erg~s$^{-1}$)}
}
\startdata
M32        & 0.085 & 25   & 3.7   & 3.1&   0.15 & 0.12 \nl
NGC~3377   & 2.0   & 580  & 88    & 5.8 &    4.7 & 5.1 \nl
M31~bulge  & 3.7   & 1100  & 160  &  2.9&    6.3 & 4.7 \nl
M105       & 4.1   & 1200  & 180  & 5.9 &   8.3 & 11 \nl
NGC~4278   & 5.5   & 1600  & 240  & 7.2 &    15 & 17 \nl
NGC~3585   & 15    &  4400  & 660  & 3.5&   38 & 23 
\enddata
\tablenotetext{a}{Table 1 of GB10}
\tablenotetext{b}{this work, Equation (\ref{sss_number_hkn})}
\tablenotetext{c}{this work}
\tablenotetext{d}{this work, Equation (\ref{total_soft_x-ray_flux})}
\end{deluxetable}










\end{document}